\documentclass[article,twocolumn,superscriptaddress,floatfix]{revtex4}

\usepackage[latin1]{inputenc}
\usepackage{bm}
\usepackage{multirow,amssymb,amsbsy,amsmath}
\usepackage{graphicx}
\usepackage{verbatim}
\usepackage{epstopdf}
\makeatletter
\usepackage{pifont}
\makeatother

\begin{document}

\title{Hyperfine Structure and Coherent Dynamics of Rare Earth Spins Explored with Electron-Nuclear Double Resonance at Sub-Kelvin Temperatures}
\author{Pei-Yun Li}
\author{Chao Liu}
\author{Zong-Quan Zhou$\footnote{email:zq\_zhou@ustc.edu.cn}$}
\author{Xiao Liu}
\author{Tao Tu}
\author{Tian-Shu Yang}
\author{Zong-Feng Li}
\author{Yu Ma}
\author{Jun Hu}
\author{Peng-Jun Liang}
\author{Xue Li}
\author{Jian-Yin Huang}
\author{Tian-Xiang Zhu}
\author{Chuan-Feng Li$\footnote{email:cfli@ustc.edu.cn}$}
\author{Guang-Can Guo}
\affiliation{CAS Key Laboratory of Quantum Information, University of Science and Technology of China, CAS, Hefei, 230026, China}
\affiliation{CAS Center for Excellence in Quantum Information and Quantum Physics,
University of Science and Technology of China, Hefei, 230026, China}
\date{\today}
\begin{abstract}
{An experimental platform of ultralow-temperature pulsed ENDOR (electron-nuclear double resonance) spectroscopy is constructed for general bulk materials. Coherent properties of the coupled electron and nuclear spins of rare-earth dopants in a crystal ($^{143}$Nd$^{3+}$:Y$_2$SiO$_5$) is investigated from 100 mK to 6 K. At the lowest working temperatures, two-pulse-echo coherence time exceeding 2 ms and 40 ms are achieved for the electron and nuclear spins, while the electronic Zeeman and hyperfine population lifetimes are more than 15 s and 10 min. With the aid of the near-unity electron spin polarization at 100 mK, the complete hyperfine level structure with 16 energy levels is measured using ENDOR technique without the assistance of the reconstructed spin Hamiltonian. These results demonstrate the suitability of the deeply cooled paramagnetic rare-earth ions for memory components aimed for quantum communication and quantum computation. The developed experimental platform is expected to be a powerful tool for paramagnetic materials from various research fields.}
\end{abstract}
\maketitle

\section{introduction}

Electron-coupled nuclear spins in solids are promising candidates for the memory components in quantum computation and quantum communication. The electron spins can be dedicated to interface with superconducting circuits \cite{xiang2013hybrid,zhu2011coherent,probst2013anisotropic}, while long-term storage is provided by the nuclear spins to access much longer coherence lifetimes \cite{ranvcic2018coherence,steger2012quantum}. Arbitrary qubit states can be coherently transferred between the electron and nuclear spins with high fidelity \cite{morton2008solid,wolfowicz2015coherent}. Large storage bandwidth is promised by the GHz-range electronic Zeeman, and electron-nuclear hyperfine interactions \cite{ortu2018simultaneous,afzelius2009multimode,saglamyurek2015quantum,ranvcic2018coherence}. This is also the appealing physical platform for the realization of optical quantum memory at the telecom wavelength with sufficient storage time and efficiency \cite{ranvcic2018coherence}.

Although there is a general interest to lower the working temperatures to decelerate the decoherence processes \cite{steger2012quantum,wolfowicz2015coherent,lim2018coherent}, coherent properties of the electron-coupled nuclear spin systems at sub-Kelvin temperatures are still not clear. As a basic tool, the coherent dynamics of the coupled electron and nuclear spins can be investigated with pulsed ENDOR spectroscopy \cite{schweiger2001principles,feher1959electron,morton2008solid}, which has only been conducted at liquid-helium temperatures for bulk materials. In this work, we achieve a comprehensive enhancement of the population and coherence lifetimes of both the electronic and the coupled nuclear spins. The measurements are based on pulsed EPR (electron paramagnetic resonance) and ENDOR spectroscopy at temperatures down to 100 mK when the electron spins are almost fully polarized. Rare-earth (RE) ion doped solids, the model system studied herein, is state-of-the-art candidate material for optical quantum memory \cite{clausen2011quantum,zhong2017nanophotonic,bussieres2014quantum,zhou2012realization,tang2015storage,hedges2010efficient} with the potential for microwave memory \cite{wolfowicz2015coherent} and microwave-to-optical quantum transduction \cite{williamson2014magneto,o2014interfacing}. For RE-ion-doped solids at sub-Kelvin temperatures, there have been previous works concerning on the optical coherence lifetimes \cite{kukharchyk2018optical,saglamyurek2015quantum,zhong2018optically}, but reports on the coherent spin dynamics are rare \cite{probst2015microwave,kindem2018characterization}. In this work as the sample temperature is reduced into the sub-Kelvin region, the growth of spin relaxation and coherence times is observed to accelerate simultaneously. In addition, with the aid of the near-unity polarization of the electron spins, the complete hyperfine structure of the ground state of $^{143}$Nd$^{3+}$:Y$_\text{2}$SiO$_\text{5}$ is experimentally resolved via a series of novel pulsed ENDOR sequences.

This paper is organized as follows: Section II describes the construction of the experimental platform and the verification of the sample temperature. Section III describes the experimental process to directly measure the ground-state hyperfine level structure using modified pulsed ENDOR technique. Section IV describes the reconstruction of the spin Hamiltonian. Section V discusses the coherent electron and nuclear spin dynamics of $^{143}$Nd$^{3+}$:Y$_\text{2}$SiO$_\text{5}$ over a large temperature scale from 100 mK to 6 K.

\begin{figure*}[tb]
\includegraphics[width=1.0\textwidth]{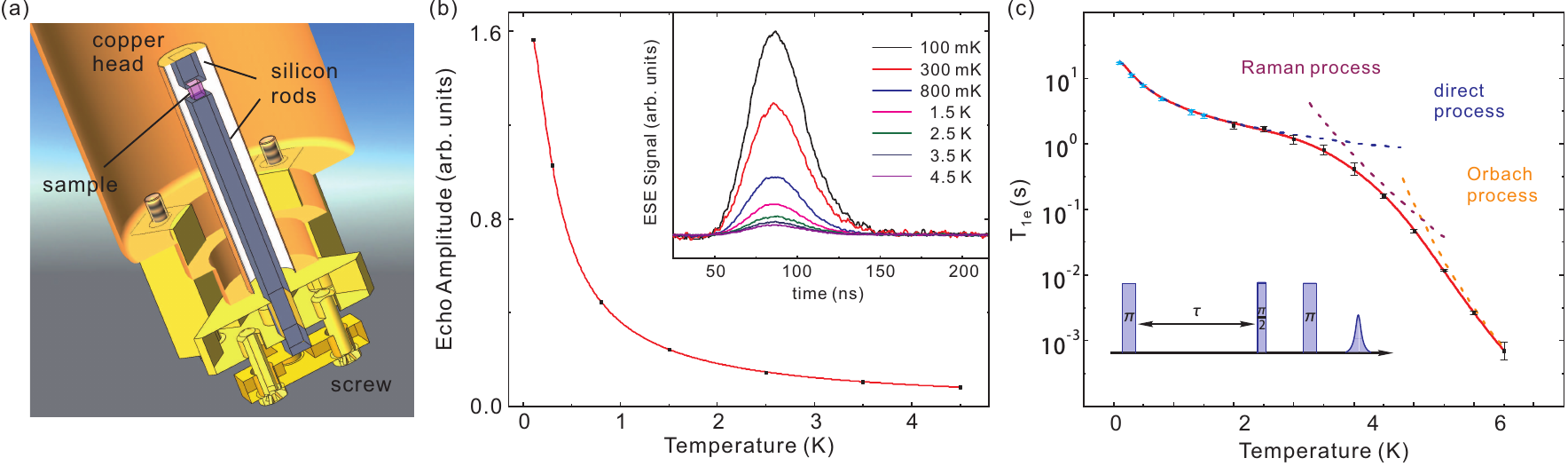}
\centering
\caption{
(a) Thermal conduction components around and inside the ENDOR resonator. Stress can be applied to the sample by the screw for better thermal contact. (b) Temperature dependence of the integrated area of ESE. The red line is the fit to a Boltzmann distribution. Profiles of the echoes recorded at various temperatures are shown in the inset. (c) Temperature dependence of $T_{1e}$ for even isotopes of Nd$^{3+}$ with $I=0$. The red solid line is the fitted curve based on the spin-lattice relaxation model and the $T_{1e}$ data above 2 K. The cyan stars represent the $T_{1e}$ measured below 2 K. The contributions from the direct, Raman and Orbach processes are also displayed independently with dashed curves colored in blue, purple and orange, respectively. }
\end{figure*}

\section{experimental setup and temperature verification}

The sample, a Y$_2$SiO$_5$ crystal doped with isotopically purified $^{143}$Nd$^{3+}$ ions of 20 ppm (Laser Crylink), is mounted inside a modified cylindrical dielectric resonator (Bruker EN4118X-MD4) with a copper head integrated with its shell above the sample space. The crystal is cut along its D1, D2 and b axes with dimensions of $1.2\times1.0\times1.4$ mm$^3$. There are some even Nd isotopes remaining in the crystal since the enrichment of $^{143}Nd$ is greater than 91$\%$ in the start material. Thermal conduction to the crystal is provided by two rods made of high-purity single-crystal silicon \cite{slack1964thermal}; see Fig. 1(a). The EPR resonance frequency is $f_r=9.56$ GHz. Cooling of the sample and the ENDOR resonator is provided by a cryogen-free dilution refrigerator (Triton 400, Oxford Instruments). The electromagnet is installed separately from the dilution refrigerator on a rotary table that is portable along a guided rail. Unless otherwise stated, the orientation of the magnetic field $B_0$ is close to the D1 optical extinction axis of the crystal (within the error of $\sim2 ^\circ$). MW and RF pulses with peak powers of 20 W and 100 W are employed to coherently drive the electron and nuclear spin ensembles, respectively. To eliminate heating induced by the background noise of the MW and RF amplifiers, TTL logic circuits are used to invalidate the amplifiers unless there are pulses injected into them. To avoid excessive Ohm heating in the coax cables, superconducting coax cables (NbTiNbTi085 and NbTiNbTi047, Keycom) are used below the 4 K stage. 0-dB attenuators are inserted to heat-sink both the inner and outer conductors.

\begin{figure}[b]
\centering
\includegraphics[width=0.32\textwidth]{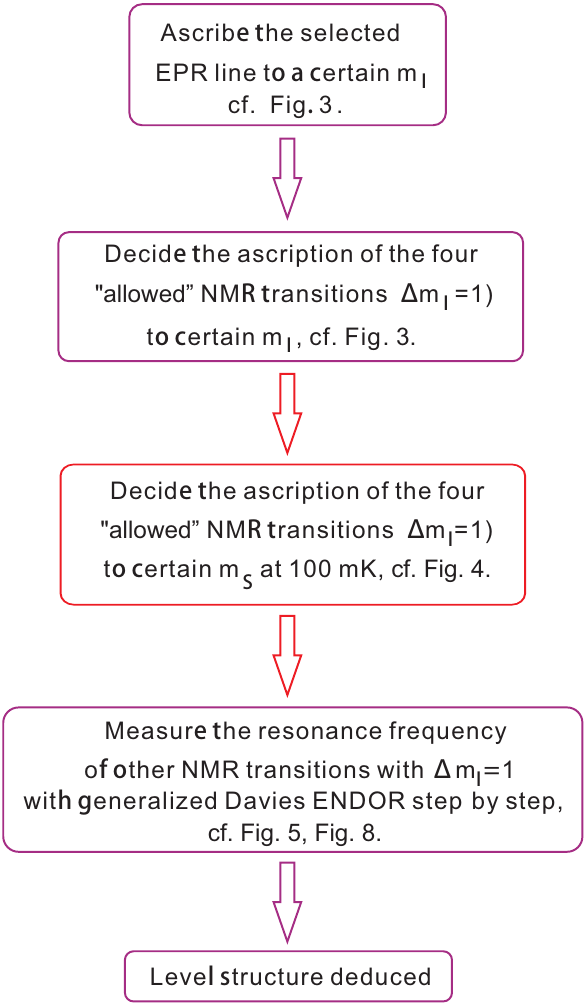}
\caption{
 A sketch to the procedures for the determination of the ground-state hyperfine level structure of $^{143}$Nd$^{3+}$:Y$_2$SiO$_5$.
}
\end{figure}

The need for high power is thought as a major challenge when performing pulsed EPR and ENDOR experiments at sub-Kelvin temperatures \cite{sigillito2017all}. Therefore, verification of the sample temperature is indispensable. Preliminarily, the integrated area of the electron spin echo (ESE), which is proportional to the electron spin polarization $P_e=\tanh(\frac{hf_r}{2k_BT})$, is used to calibrate the sample temperature. Here, $h$ is the Planck constant, $k_B$ is the Boltzmann constant, and $T$ represents the sample temperature. To take advantage of the simple level structure of a standard spin-1/2 system, $B_0$ is set to 458.2 mT. The remaining even isotopes of Nd$^{3+}$ with $I=0$ are addressed (see Fig. 2(a)). ESE is produced by conventional two-pulse-echo sequences. The $\pi$ pulses are 48 ns in length and $\tau$ is set to 1 $\mu$s, much shorter than the electron spin coherence time. It can be observed from Fig. 1(b) that the integrated area of ESE at various temperatures agrees very well with the fitted line. At 100 mK, the electron spin polarization of $98\%$ is achieved. In this regime the cross relaxation among the electron spins can be negligible \cite{cruzeiro2017spectral,kutter1995electron}.

\begin{figure}[tb]
\centering
\includegraphics[width=0.35\textwidth]{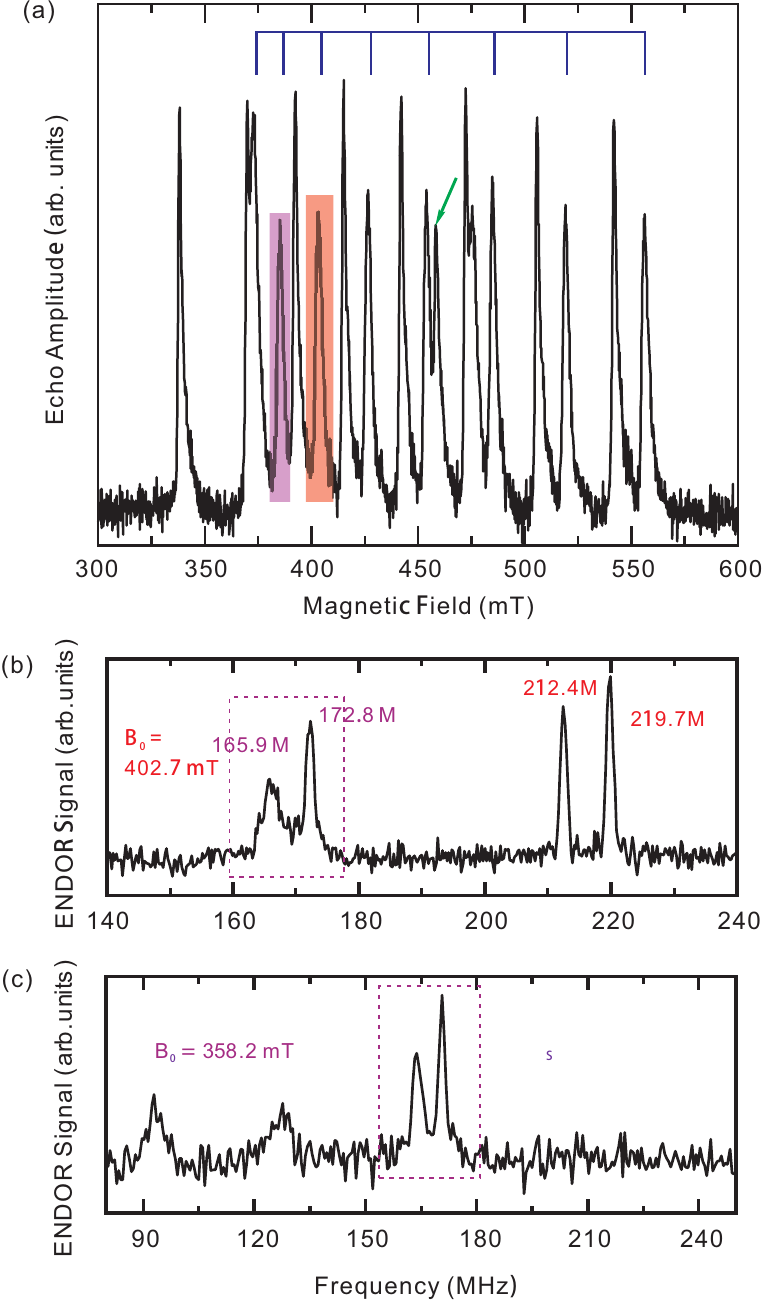}
\caption{
(a) Result of the Field-swept ESE experiment. The resonance line with $B_0=458.2$ mT used for temperature verification is marked with a green arrow. The hyperfine level structure is determined at the EPR line with $B_0=402.7$ mT, which is shaded in blue. The EPR line at 358.2 mT is shaded in purple. (b)(c) Davies ENDOR spectrum recorded at $B_0=402.7$ mT and 358.2 mT, respectively.
}
\end{figure}

\begin{figure}[b]
\centering
\includegraphics[width=0.32\textwidth]{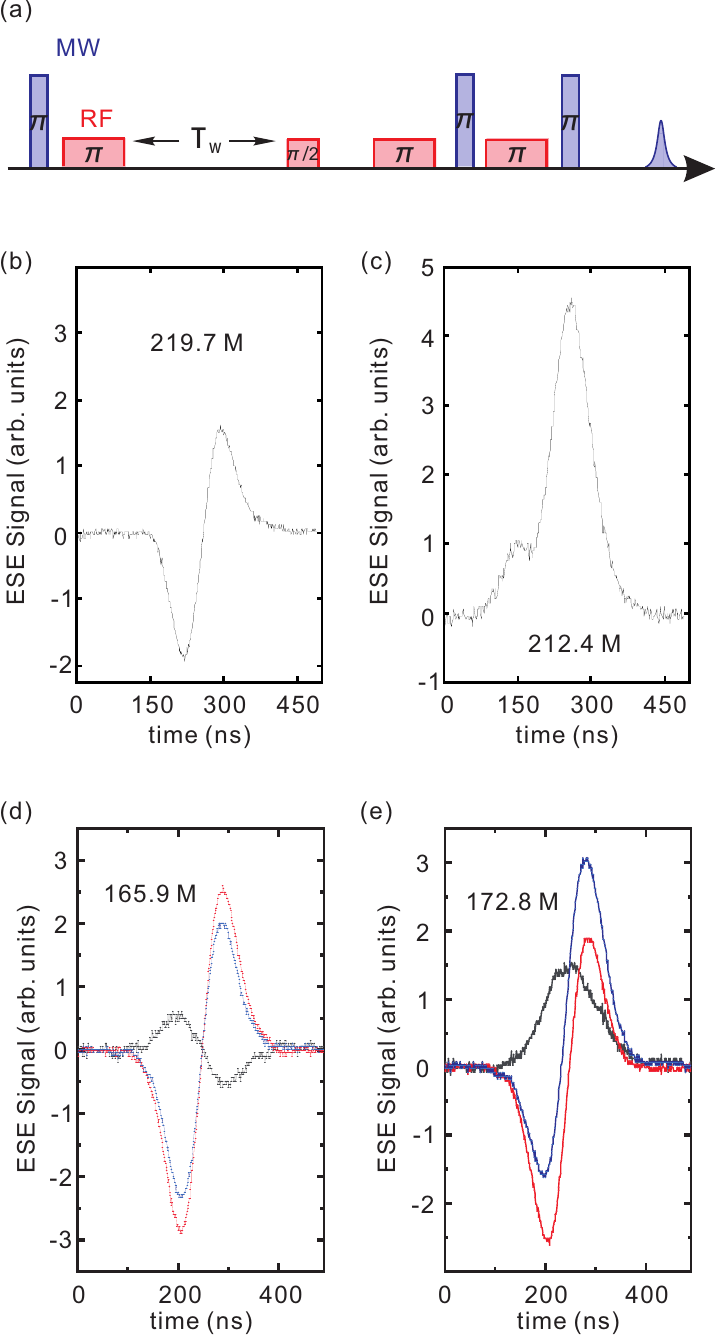}
\caption{
(a) Pulse sequence applied to determine the correspondence of the NMR lines to the electron spin projection $m_S$. This sequence can also be used to measure $T_{1n}$ by varying the length of $T_W$ while $T_{1e}\ll T_W\ll T_{1n}$ is satisfied. (b)-(e) Spin-echo signal recorded at the end of the sequence presented in (a) when RF is set to 219.7 MHz, 212.4 MHz, 165.9 MHz, and 172.8 MHz, respectively. In (d) and (e), the echo signals recorded with (shown in blue) and without (shown in red) the initial nuclear spin polarization, and the difference between them (shown in grey), are presented.
}
\end{figure}

The sample temperature is also verified by means of the dynamical process of the electron spins. The electron spin relaxation time $T_{1e}$ is measured with inversion recovery sequences. Its temperature dependence can be modelled with the well-established spin lattice relaxation (SLR) mechanism which is detailed in Sec. IV. We use the $T_{1e}$ data at higher temperatures from 2 K to 6.5 K to generate the fitted curve in Fig. 1(c), and extrapolate it down to 100 mK.  $T_{1e}$ recorded at lower temperatures perfectly follows this predicted curve. The longest $T_{1e}$ of 17 s is obtained at 100 mK, which has reached the limit imposed by the spontaneous emission of phonons \cite{jeffries1963dynamic,supp}.

\section{direct measurement of hyperfine level structure}

Taking advantage of the ultra-low sample temperature, here we develop a series of novel ENDOR sequences to directly measure the complete ground-state hyperfine level structure of $^{143}$Nd$^{3+}$:Y$_2$SiO$_5$. The method is free from additional optical transitions \cite{ranvcic2018coherence,ortu2018simultaneous}, or the reconstructed spin Hamiltonians \cite{weil2007electron,guillot2006hyperfine}, as required in previous studies. A guideline is given in Fig. 2.

The field-swept ESE spectrum is given in Fig. 3(a). It is recorded at 6.0 K with $B_0$ close to the D1 axis.  The 18 resonance lines come from Nd$^{3+}$ located in two magnetically inequivalent classes that are related by a C$_2$ symmetry along the crystal's b axis \cite{wolfowicz2015coherent}. Each of the classes contains 8 lines corresponding to $^{143}$Nd$^{3+}$ with the nuclear spin of $I=7/2$, and one other line corresponding to even isotopes with $I=0$. In this study we try to characterize the hyperfine level structure at B$_0$ = 402.7 mT, when one of the EPR transitions is at resonance.

\begin{figure}[t]
\centering
\includegraphics[width=0.36\textwidth]{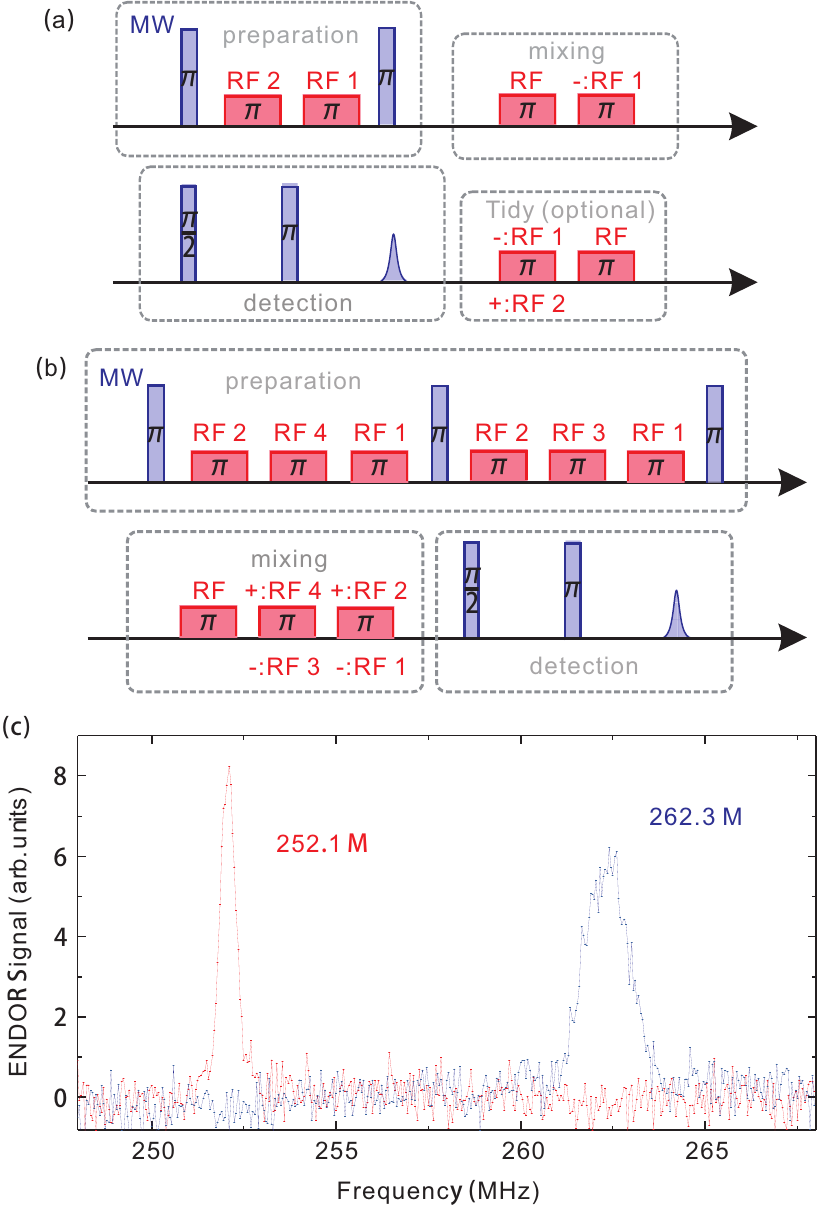}
\caption{
(a) Pulse sequence used to search for the NMR transitions $|m_S,-1/2\rangle\to|m_S,+1/2\rangle$. (b)Pulse sequence used to search for the NMR transitions $|m_S,-3/2\rangle\to|m_S,-1/2\rangle$. Here $m_S$ can be either -1/2 (to choose ``-'' in the fig
ures) or +1/2 (to choose ``+'' in the figures). RF1, RF2, RF3, and RF4 correspond to the excitation frequencies of 212.4 MHz, 219.7 MHz, 172.8 MHz, and 165.9 MHz, respectively. RF represents the frequency to be swept. (c) ENDOR signals corresponding to $|-1/2,-1/2\rangle\to|-1/2,+1/2\rangle$ (shown in red) and $|+1/2,-1/2\rangle\to|+1/2,+1/2\rangle$ (shown in blue).}
\end{figure}

For a selected EPR transition, the ``allowed" NMR transitions with $\Delta m_S=0$ and $\Delta m_I=1$ can be detected by Davies ENDOR sequences \cite{davies1974new}. Here we use the ket $|m_S,m_I\rangle$ to denote the electron and nuclear spin projections. The ENDOR spectra measured at 402.7 mT and 358.2 mT are shown in Fig. 3(b) and Fig. 3(c), respectively. It can be seen that the two resonance peaks located at $\sim$ 165 MHz and 172 MHz coexist in the two spectra, indicating that these two lines correspond to the $^{143}$Nd$^{3+}$ ions located in the same magnetically inequivalent class. By comparing Davies ENDOR results acquired at various resonant magnetic fields, all of the 8 EPR lines corresponding to the same magnetically inequivalent class can be accordingly discriminated, as marked in Fig. 3(a). Consequently, it can be deduced that the 402.7-mT line corresponds to the EPR transition of $|-1/2,+3/2\rangle\to|+1/2,+3/2\rangle$. Two ENDOR lines located at 212.4 MHz and 219.7 MHz, as shown in Fig. 2(b), correspond to the NMR transitions of $|m_S,+1/2\rangle\to|m_S,+3/2\rangle$, in which $m_S$ can be +1/2 or -1/2.

A critical step is to decide the electron spin projection $m_S$ for these lines \cite{wolfowicz2015coherent,lim2018coherent,supp}. This can be experimentally resolved at 100 mK when the electron spin is almost fully polarized. As shown in Fig. 4(a), nuclear spin polarization can then be established after the MW and RF $\pi$ pulses and the subsequent waiting time of $T_W=60$ s. If the RF frequency is resonant with the NMR transition in the lower electron spin state ($m_S=-1/2$), nuclear spin coherence can be created by an RF $\pi /2$ pulse and transferred to the electron spin \cite{morton2008solid}. Finally, an ESE signal can be observed. Otherwise, operations starting at the emptied energy level with $m_S=+1/2$ are ineffective. An illustration to the pulse sequence is given in Fig. S1 in the supplemental material (SM) \cite{supp}. The experimental results are shown in Fig. 4(b)-(e). The experimental phenomena are less clear for the NMR transitions of $|\pm1/2,+3/2\rangle\to|\pm1/2,+5/2\rangle$ at 165.9 MHz and 172.8 MHz, possibly due to the imperfect RF excitation or the electron-nuclear cross relaxation. Nevertheless, the signal recorded without the initial nuclear spin polarization can be subtracted from the signal recorded at the end of the entire pulse sequence to acquire a standard echo profile. The experimental results indicate that the 212.4-MHz and 172.8-MHz NMR transitions are in the lower electron spin level, corresponding to $m_S=-1/2$, while the 219.7-MHz and 165.9-MHz NMR transitions are in the upper electron spin level, corresponding to $m_S=+1/2$.

\begin{figure}[tb]
\centering
\includegraphics[width=0.22\textwidth]{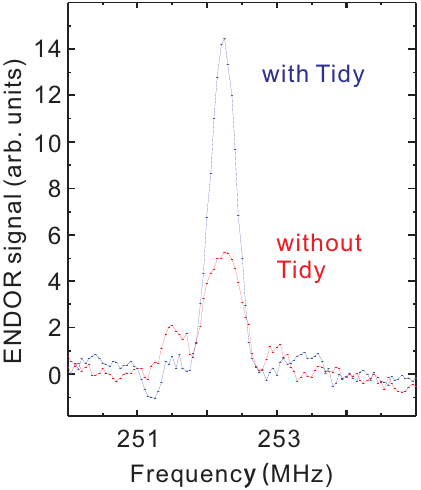}
\caption{
The ENDOR signal at 252.1 MHz measured with and without the ``Tidy'' sequence. The ENDOR signal is acquired at approximately 6 K. The experimental repetition time is 20 ms, and the average number is set as 50.}
\end{figure}

\begin{figure}[b]
\centering
\includegraphics[width=0.36\textwidth]{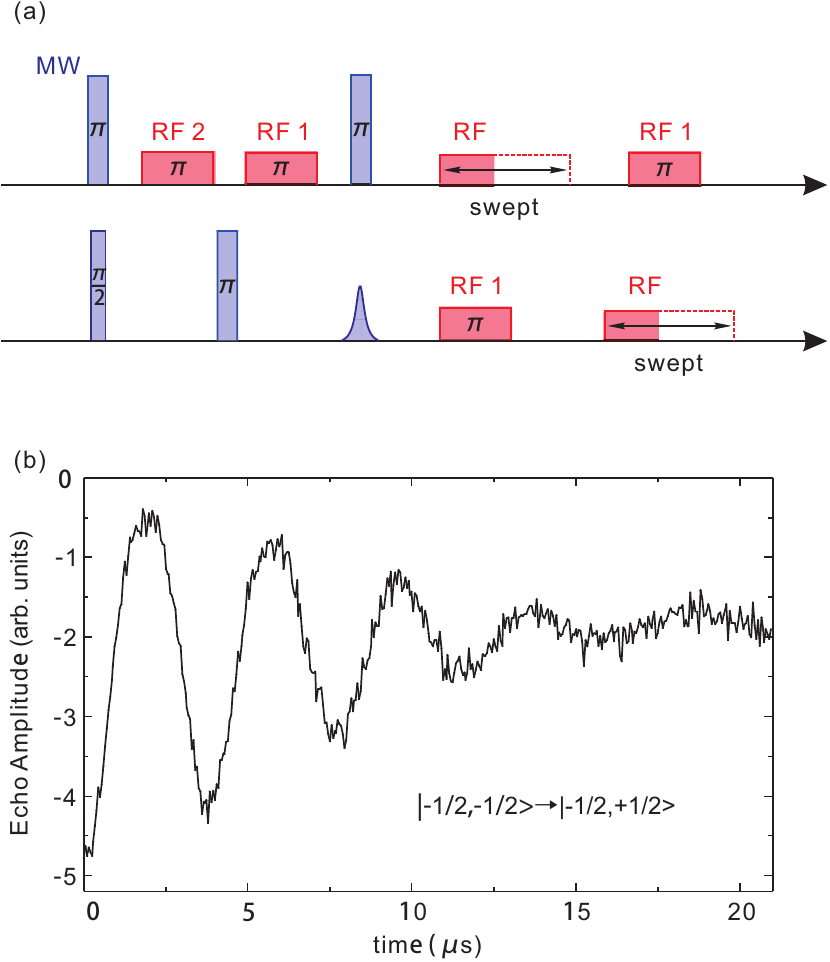}
\caption{
(a) The pulse sequence applied to detect the Rabi oscillation of the NMR transition $|\pm1/2,-1/2\rangle\to|\pm1/2,+1/2\rangle$ with its experimental result given in (b).}
\end{figure}

To acquire the complete hyperfine level structure for $^{143}$Nd$^{3+}$ with $I=7/2$, the NMR transitions which are unreachable by standard pulsed ENDOR experiments are further investigated using a generalized Davies ENDOR technique. Single spin flips during the preparation and mixing steps in a traditional Davies ENDOR sequence \cite{davies1974new,schweiger2001principles} can be revised into successive multiple spin flips in order to access NMR transitions that do not share a common energy level with the selected EPR transition. These transitions of $|m_S,m_I\rangle\to|m_S,m_I+1\rangle$ with $m_I<+1/2$ ($m_I>+3/2$) can be detected in a step-by-step manner once the resonance frequencies of $|m_S,m_I^\prime\rangle\to|m_S,m_I^\prime+1\rangle$, with $m_I<m_I^\prime\leq+1/2$ ($m_I>m_I^\prime\geq+3/2$), have been determined. Two sequences are given in Fig. 5 as examples. A graphical illustration is given in Fig. S2 in the SM \cite{supp}. In the preparation step, the population is pumped from $|-1/2,m_I^\prime\rangle$ to $|+1/2,m_I-m_I^\prime+1/2\rangle$ for $m_I^\prime=+3/2, -1/2,..., m_I+2$, and $m_I+1$, successively. In the mixing step, in which the RF frequency is swept, the EPR transition is depolarized by the population transfer from $|m_S,m_I\rangle$, in which $m_S$ can be deliberately chosen to be either +1/2 or -1/2. In the detection step, the spin echo intensity is monitored to produce the ENDOR spectra. The ``Tidy" pulse can also be generalized to the pulse sequence. By compulsively initializing the population distribution among the hyperfine energy levels, the reduction of the ENDOR signal intensity due to the slow nuclear spin relaxation can be effectively avoided \cite{tyryshkin2006davies}.

In Fig. 5(c), two ENDOR lines of the NMR transitions $|\pm1/2,-1/2\rangle\to|\pm1/2,+1/2\rangle$ are given. To note, in principle the specific ENDOR line corresponding to $m_S=+1/2$ or $-1/2$ should appear in only one of the two spectra. One specific ENDOR sequence is designed for one specific NMR transition. This is different from standard Davies ENDOR experiments. The analysis of the correspondence of a particular ENDOR line to the certain electron spin projection $m_S$ is therefore simplified.

The effect of the generalized ``Tidy" sequence is shown in Fig. 6. In addition, with the variation of the RF pulse length as illustrated in Fig. 7(a), the Rabi oscillation of an NMR transition that is unreachable with the conventional pulsed ENDOR technique can be observed. The result is given in Fig. 7(b), demonstrating the ability to coherently manipulate all of the NMR transitions with $\Delta m_I=1$.

\begin{figure}[t]
\centering
\includegraphics[width=0.36\textwidth]{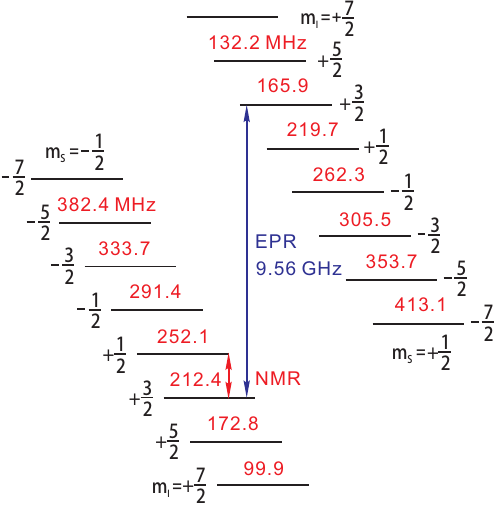}
\caption{
Complete energy level structure of the ground state $^4$I$_{9/2}$(0) of $^{143}$Nd$^{3+}$:Y$_{2}$SiO$_{5}$, as determined with the pulsed ENDOR technique.}
\end{figure}

All other relevant ENDOR signals corresponding to the transitions with $\Delta m_S=0$, $\Delta m_I=1$ are shown in Fig. S3 in the SM \cite{supp}. After gathering the results from these ENDOR signals, a complete energy level diagram can be generated, which is given in Fig. 8.

\section{reconstruction of the spin hamiltonian}

\begin{figure}[bt]
\centering
\includegraphics[width=0.3\textwidth]{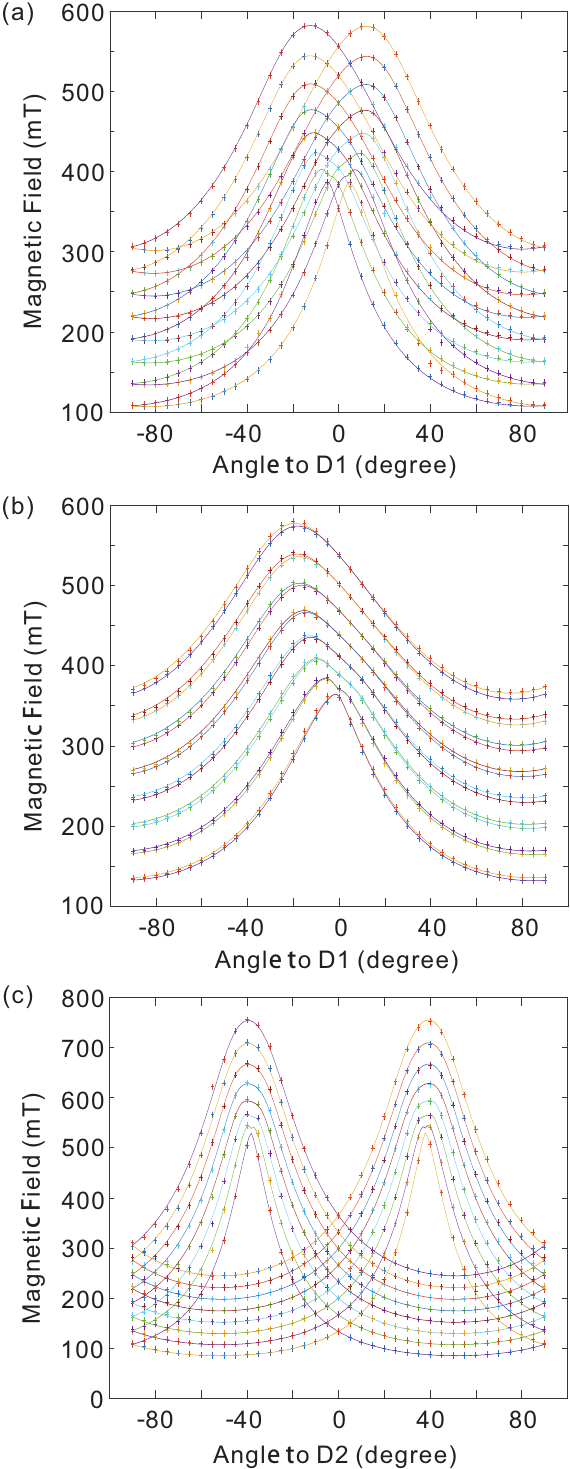}
\caption{
Angular variations of the EPR peaks of $^{143}$Nd$^{3+}$:Y$_2$SiO$_5$. The magnetic field B$_0$ varies on the bD1 (a), D1D2 (b), and bD2 (c) planes. The fitted results are represented by solid lines.}
\end{figure}

To make a comparison, we reconstructed the spin Hamiltonian to predict the hyperfine level structure. This Hamiltonian is also useful in the analysis of the decoherence mechanisms in the Section V. The spin Hamiltonian takes the form as:
\begin{equation}
  H=\beta \mathbf{B_0}\cdot\mathbf{g}\cdot\mathbf{S}+\mathbf{I}\cdot\mathbf{A}\cdot\mathbf{S}
\end{equation}
where $\beta$ is the Bohr magneton; $\mathbf{g}$ and $\mathbf{A}$ are the g and hyperfine tensors, respectively; and $\mathbf{S}$ ($\mathbf{I}$) denotes the electron (nuclear) spin operator.

Three pieces of crystals with dimensions of 13 mm$\times$2.1 mm$\times$2.1 mm are used here. They are cut along the D1, D2 and b axes, while the 13-mm orientation is different for each of the crystals. Field-swept ESE experiments are performed when the crystals are rotated around the 13-mm orientation in increments of 5 degrees. We note that Nd$^{3+}$ ions can substitute for Y$^{3+}$ ions located at two of the crystallographic sites, referred to as site I and site II \cite{beach1990optical}. However, in most cases, only the EPR lines from site I can be completely gathered, possibly because the Nd$^{3+}$ dopants preferentially occupy site I \cite{wolfowicz2015coherent}. When ${B_0}$ has an arbitrary orientation with respect to the crystal (i.e. not parallel or perpendicular to the b axis), each of the sites can be divided into two magnetically inequivalent classes. This can lead to two sets of EPR signals as displayed in each of the subfigures of Fig. 9. Nevertheless, during the fitting process for each of the crystal orientation the resonant magnetic field positions are recorded in order from small to large, regardless of the different magnetic inequivalent sites.

A program based on the EasySpin software package \cite{stoll2006easyspin} is used for the fitting process. When a group of g and A matrices is generated by the program, the peak positions are calculated and arranged also in order from small to large. These calculated results are then compared with the experimental data. The assignment of the peaks to the particularly predicted lines is not needed. Ultimately, an average deviation of $18.4$ Gauss for a single experimental data point is achieved. The best-fit $\mathbf{g}$ matrix in the D2bD1 crystal frame is:
$$g=\left(
  \begin{array}{ccc}
    -1.03 & -2.48 & 0.44 \\
    -2.49 & -2.19 & -0.14 \\
    0.44 & -0.14 & 1.39 \\
  \end{array}
\right)
$$

The principal values are $g_x=-4.16$, $g_y=0.68$, and $g_z=1.65$. The corresponding Euler angles ($ZYZ$ convention) are $\alpha=87^{\circ}$, $\beta=149^{\circ}$, and $\theta=36^{\circ}$, which describe the transformation from the principal axis frame to the D2bD1 crystal frame.

The best-fit $\mathbf{A}$ matrix in the D2bD1 frame is (in MHz):
$$A=\left(
  \begin{array}{ccc}
    495.7  &687.4 &-232.8\\
        		687.4 & 751.8 & 165.8\\
       		-232.8  &165.8 &-338.3\\
  \end{array}
\right)
$$
The principal values are $A_x=1323$, $A_y=-520$, and $A_z=-137$ MHz, and the corresponding Euler angles ($ZYZ$ convention) are $\alpha=91^{\circ}$, $\beta=122^{\circ}$, and $\theta=-140^{\circ}$.

Based on the reconstructed spin Hamiltonian, simulation shows that when the EPR spectrum in Fig. 3(a) is generated, using the language of the spherical coordinate system, the orientation of B$_0$ is $(\theta,\phi)=(-2.24^{\circ},-66.35^{\circ})$ with respect to (D2, b, D1).

We note that the simulated level structure at $B_0=402.7$ mT using the spin Hamiltonian coincident with our measured level structure with deviations from $\sim1$ MHz to $\sim30$ MHz. The advantage on the accuracy of our modified ENDOR technique is exhibited.

\section{coherent electron and nuclear spin dynamics}

\begin{figure}[t]
\centering
\includegraphics[width=0.36\textwidth]{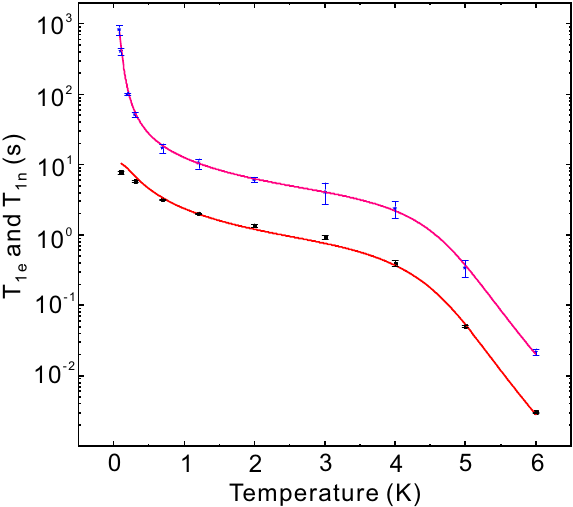}
\caption{
Temperature dependence of the electron and nuclear spin relaxation times recorded at $B_0=402.7$ mT.
The measured $T_{1n}$ values are shown in blue squares. The NMR transition $|-1/2,+1/2\rangle\to|-1/2,+3/2\rangle$ (212.4 MHz) is investigated here. The $T_{1e}$ data are shown in black squares. The solid lines are the corresponding fitted curves.}
\end{figure}

\begin{figure}[bt]
\centering
\includegraphics[width=0.5\textwidth]{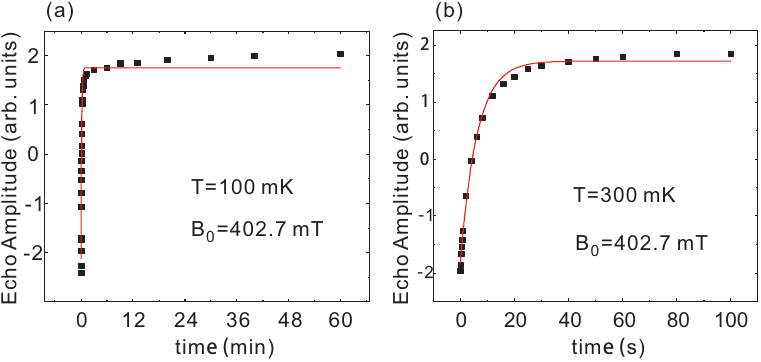}
\caption{
Inversion-recovery curves measured for $^{143}$Nd$^{3+}$:Y$_2$SiO$_5$ at temperatures lower than 500 mK. Solid lines are single-exponential fits.}
\end{figure}

A comparison of $T_{1e}$ and the nuclear spin relaxation time $T_{1n}$ recorded at B$_0$ = 402.7 mT is given in Fig. 10. $T_{1e}$ is measured with inversion recovery experiments. $T_{1n}$ is measured using the sequence presented in Fig. 4(a). The MW and RF $\pi$ pulses are 52 ns and 1.08 $\mu$s in length. For a coupled electron-nuclear spin system, the relaxation route can be more complicated compared with a simple spin-1/2 system. After the $\pi$-pulse excitation at the beginning of the inversion-recovery sequence, the population can, as usual, drop directly to the energy level with the same nuclear spin projection $m_I$; while dropping firstly to levels with different $m_I$, and then to take a long time coming back via nuclear spin relaxation, can also be the possible relaxation routes. It is observed that when the sample temperature is lower than 500 mK, single-exponential fits cannot work well, see Fig. 11. However when the temperature is higher than 500 mK, single-exponential fits are pretty good. The significant change happened below the electronic Zeeman temperature seems to be due to the rapid increase of the ratio between the electron and nuclear spin relaxation rates, but further investigation is still required. For integrity, the electron spin relaxation times displayed in Fig. 10 are still the results of single-exponential fits over the full temperature range.

For $^{143}$Nd$^{3+}$, nuclear spin relaxation is dominated by the rapid relaxation of the electron spins at temperatures above the electronic Zeeman temperature of $hf_r/k_B\sim460$ mK. This behavior manifests as a constant ratio $\sigma$ between the two relaxation rates \cite{jeffries1963dynamic}, which is obvious in Fig. 10. However, as the temperature decreases, the electron spins start to freeze out, and the much slower SLR of the nuclear spins themselves starts to take the place. The nuclear spin relaxation can thus be modelled as follows \cite{waugh1988mechanism,abragam2012electron}:
\begin{equation}
\begin{split}
T_{1n}^{-1}=\sigma (1-P_e^2) T_{1e}^{-1} + \gamma_d \coth(\frac{hf_n}{2k_B T}) + \gamma_r T^9 \\
+ \gamma_o f_r^3\exp(-\frac{hf_r}{k_{B}T}),
\end{split}
\end{equation}
where $f_n$ represents the NMR transition frequency of 212.4 MHz, and
$\gamma_d$, $\gamma_r$, and $\gamma_o$ are the coupling factors of the direct, Raman and Orbach processes, respectively. The ratio $\sigma$ is fitted to be 0.126. The coupling factors for direct and Orbach processes are fitted as $\gamma_d=7.3\times10^{-5}$ s$^{-1}$ and $\gamma_o=1\times10^{-32}$ Hz$^{-2}$; while the impact of the Raman process is found to be negligible. The $T_{1n}$ data show excellent agreement with the fitted curve. Moreover, there is a strong temperature dependence for $T_{1n}$ even below 100 mK. $T_{1n}$ of $13.8\pm2.3$ min is obtained at $75$ mK, which is much longer compared with the optically excited state lifetime of $\sim300$ $\mu$s \cite{usmani2010mapping}. This is highly profitable for the efficient optical pumping and long-term quantum storage using $^{143}$Nd$^{3+}$:Y$_2$SiO$_5$ \cite{ranvcic2018coherence,tang2015storage}.

\begin{figure}[bt]
\centering
\includegraphics[width=0.38\textwidth]{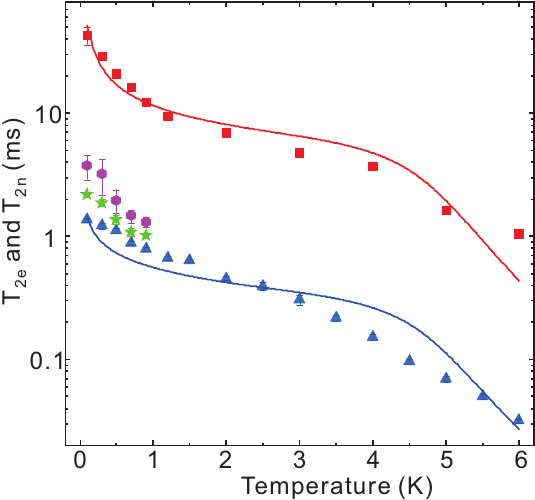}
\caption{Temperature dependence of the electron and nuclear spin coherence times recorded at $B_0=402.7$ mT.
The measured $T_{2n}$ values are shown in red squares along with the red solid line as the fitted curve. $T_{2e}$ values measured with soft pulses and the nominal flipping angles of $\pi$ are shown in blue triangles. The blue solid line is the fitted curve. The green stars are $T_{2e}$ data measured with the refocusing pulses of 200 ns. These measurements are conducted below 1 K for a high single-to-noise ratio. The purple hexagons represent the deduced $T_{2e}$ values, from which the ID effect has been eliminated.}
\end{figure}

The study of $T_{2e}$ is conducted during the second period of cooling, utilizing the same EPR transition. The temperature dependence of $T_{2e}$ is given in Fig. 12 along with the fitted curve based on the model discussed in the SM \cite{supp} in detail. Spectral diffusion (SD) caused by electronic SLR-induced random spin flips is the main source of decoherence for both the electron and nuclear spins. The decay of spin echoes therefore follows the Mims decay law \cite{mims1968phase} taking the form as $\exp[-(2\tau/T_2)^m]$. In Y$_2$SiO$_5$ the Nd$^{3+}$ ions preferentially occupy one of the Y$^{3+}$ crystallographic sites \cite{wolfowicz2015coherent}, but as shown in Fig. S4 of SM \cite{supp}, the impact of the dopant ions located at the other site should not be omitted. This can lead to inaccuracy to the modelling. When the temperature is reduced, the SLR rate would slow down, and the electron spins can be eventually frozen out. As a result, the electronic SLR-induced decoherence effect can be gradually inhibited. In order to alleviate the instantaneous diffusion (ID) effect which results from the compulsory electron spin flips following the refocusing pulse in a two-pulse Hahn echo sequence \cite{klauder1962spectral}, here soft pulses are utilized in the $T_{2e}$ measurements \cite{dzuba1996selective}. The MW peak power is set to 75 mW and the nominal $\pi$ pulses are 700 ns. At 100 mK, $T_{2e}$ of 2.18$\pm$0.09 ms is obtained with a reduced refocusing pulse of 200 ns. The stretch factor is $m=1.42\pm0.09$. This is a remarkable result obtained for RE ions without the assistance of a clock transition, and it have been comparable to other solid-state electron spin systems such as defects in diamonds and in SiC.

The temperature dependence of $T_{2n}$ is also shown in Fig. 12 along with the fitted curve based on Eq. (5) in the SM \cite{supp}. $T_{2n}$ is measured by converting nuclear spin echoes into ESE \cite{morton2008solid}. $T_{2n}$ of $43\pm7$ ms with a stretch factor of $m=1.65\pm0.24$ is obtained at 100 mK. Once the electron spins have been frozen out, $T_{2n}$ is ultimately limited by the flip-flops of the host nuclear spins. From the effective spin Hamiltonian, see the SM \cite{supp}, the gradient of the NMR transition frequency with respect to the external magnetic field can be calculated as $160$ MHz/T. Considering the flip-flops among the bulk $^{89}$Y spins far from the central Nd$^{3+}$ ions, $T_{2n}$ can be accordingly estimated to be $\sim$ 250 $\mu$s. However, the experimentally measured $T_{2n}$ is more than 2 orders of magnitude longer. This is the result caused by the frozen core effect induced by the large magnetic moment of the central Nd$^{3+}$ electron spins \cite{zhong2015optically,ranvcic2018coherence}. Strong hyperfine coupling between the central electron spin and each of the neighbouring host nuclear spins can cause detuning among the host nuclear spins. The resonant flip-flops can thus be heavily suppressed.

\begin{figure}[bt]
\centering
\includegraphics[width=0.37\textwidth]{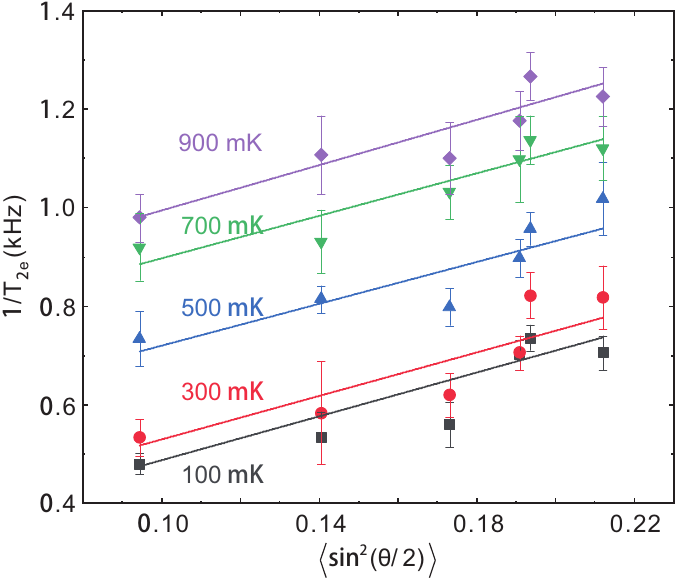}
\caption{ A demonstration of the instantaneous diffusion effect. $1/T_{2e}$ is plotted as a function of the average spin flip probability $\langle\sin^2(\theta/2)\rangle$. $T_{2e}$ is measured at B$_0$ = 402.7 mT using the pulse sequence of $\pi/2-\tau-\theta-\tau-echo$, in which $\theta$ represents the effective rotation angle of the refocusing pulse. The value of $\langle\sin^2(\theta/2)\rangle$ is calculated with Eq. (4) in the SM \cite{supp}. The MW power is set to approximately 75 mW and the refocusing pulse lengths are varied. Experiments are conducted at 5 different temperatures below 1 K. Solid lines are the linear fits.}
\end{figure}

In order to further extend the coherence lifetime of the electron spin, besides the impacts of the host nuclear spins, at least two limiting factors are needed to be considered. The first one is the remaining ID effect. Under the reduced refocusing pulse of 200 ns, the ID effect can still account for a $T_{2e,ID}=5.1$ ms as indicated in Fig. 13. Secondly, When measuring $T_{2e}$ in an electron spin ensemble, all of the resonant electron spins are excited, and these electron spins are no longer polarized. At 100 mK this would cause a $T_{2e,SD}\approx10.5$ ms. These two effects can be minimized in a low-doping sample, or even better, in single-ion systems \cite{zhong2018optically}. To overcome the superhyperfine limit for both the dopant electron and nuclear spins, in principle the most efficient way is to adopt a host material with much lower nuclear spin concentration \cite{tyryshkin2012electron,phenicie2019narrow}. Otherwise, the detrimental effect of the host nuclear spins should be somehow inhibited. For example, the zero-order Zeeman technique and dynamical decoupling can be applied for both of the MW and RF transitions \cite{zhong2015optically,ortu2018simultaneous}.

\section{summary and conclusion}

In conclusion, comprehensive enhancement of the population and coherence lifetimes is achieved in the sub-Kelvin temperature regime for both the electron and nuclear spins, with the compatibility of high-quality and fast manipulations \cite{supp}. The pulsed ENDOR protocol used for hyperfine structure characterization can also be generalized to other spin systems for which the spin Hamiltonians are hard to be precisely determined \cite{saglamyurek2015quantum,chen2018hyperfine}, and those without an auxiliary optical transition.

In addition, the compatibility of the sub-Kelvin sample temperature with a conventional 3D ENDOR resonator makes the established platform a versatile equipment suitable for ordinary EPR samples from various research fields.
For example, it is also promising to enhance the coherent properties of molecular nanomagnets \cite{melhuish2017sub,shiddiq2016enhancing}. More intrinsically, the duration of a pulsed EPR experiment should be less than the spin relaxation time, which is typically highly dependent on the sample temperature. Therefore with a wider working temperature range, broader options of the samples and the experimental pulse sequences can be expected. For instance, the rapid increase of the nuclear spin relaxation time at sub-Kelvin temperatures, as observed in this study, is possible to provide some previously inaccessible ENDOR signals for the [NiFe]-hydrogenase \cite{foerster2005orientation}, and for the Mn cluster in Photosystem II \cite{jin2014electronic}. Besides, a high degree of electron spin polarization at sub-Kelvin temperatures can also lead to an opportunity for the ground-spin-state investigation \cite{yang2012observing,bertaina2008quantum}, and a complete characterization to the hyperfine coupling constants \cite{baute2004carboxylate}.

\bibliographystyle{apsrev4-2}

\bibliography{ref}

{\bf  Acknowledgments}
This work was supported by the National Key R$\&$D Program of China (No. 2017YFA0304100), the National Natural Science Foundation of China (Nos. 11774331,11774335,11504362,11821404,11654002), Anhui Initiative in Quantum Information Technologies (No. AHY020100), Key Research Program of Frontier Sciences, CAS (No. QYZDY-SSW-SLH003), Science Foundation of the CAS (No. ZDRW-XH-2019-1), the Fundamental Research Funds for the Central Universities (No. WK2470000026).

\end{document}